\documentclass[preprint,aps]{revtex4-2}
\usepackage{color}
\usepackage{mathtools}
\usepackage{array}
\usepackage{tabularx}
\usepackage{diagbox}
\usepackage{tikz}
\usetikzlibrary{decorations.pathmorphing}
\usepackage{amsmath,bm,amssymb,amsfonts,dcolumn,color,graphicx,latexsym,epsfig}
\usepackage{amsmath,amsfonts,amsthm}
\usepackage{rsfso}
\usepackage{hyperref}
\hypersetup{
colorlinks=true,
linkcolor=magenta,
filecolor=magenta,      
urlcolor=blue,
citecolor=blue,
}
\usepackage{multirow}
\usepackage{natbib}
\usepackage{float}
\usepackage{booktabs}
\usepackage{pifont}
\usepackage{mathrsfs}
\usepackage{caption}
\usepackage{caption, threeparttable}
\begin{document}

\title{Aspects of a novel nonlinear electrodynamics in flat spacetime and in a gravity-coupled scenario}

\author{Anjan Kar}
\email{anjankar.phys@gmail.com}
\affiliation{Department of Physics, Indian Institute of Technology, Kharagpur 721 302, India}

\begin{abstract}
\noindent A novel nonlinear electrodynamics (NLE) model with two dimensionful parameters is introduced and investigated. Our model obeys the Maxwellian limit and exhibits behaviour similar to the Born-Infeld Lagrangian in the weak field limit. It is shown that the electric field of a point charge in this model has a definite maximum value. Thus, the self-energy of the point charge is finite. The phenomenon of vacuum birefringence is found to occur in the presence of an external uniform electric field. Causality and unitarity conditions for all background electric fields hold, whereas, for magnetic fields, a restricted domain of validity is found. Moreover, a minimal coupling of Einstein's General Relativity (GR) with this NLE results in solutions of regular black holes or naked singularities, depending on whether the source is a nonlinear magnetic monopole or an electric charge, respectively.
\end{abstract}


\maketitle

\newpage

\section{Introduction}\label{I}
\noindent In Maxwell's electrodynamics, the electric field and self-energy of a point charge exhibit a divergence as the radial distance approaches zero. One of the well-known procedures to address this issue involves modifying Maxwell's linear Lagrangian in a nonlinear manner. In the 1930s, Born and Infeld first developed their nonlinear model, largely known as Born-Infeld (BI) electrodynamics \cite{Born1, Born2}. The classical BI Lagrangian imposes a maximum limit on the intensity of the electric field, preventing it from becoming infinite at small distances. As a result, the energy of a point electric charge at its location becomes finite. Subsequently, several other exciting nonlinear modifications of Maxwell's Lagrangian have emerged in the literature, collectively referred to as `nonlinear electrodynamics' (NLE). For example, the effective Euler-Heisenberg Lagrangian \cite{Heisenberg} originating from the 1-loop QED calculation and the one-parameter family of ModMax NLE theory \cite{Townsend} are worth mentioning as viable alternatives. Recently, there have been various advancements in constructing novel NLE models, all of which aim to resolve the issue of singularity in the linear model \cite{Soleng, Kruglov1, Kruglov2, Kruglov3, Kruglov6, Kruglov7, Gaete, Hendi, Gullu, Balart, Mazharimousav}, with notable features. Each of them can be identified by a parameter $\eta$. When $\eta\to 0$ or $\eta\to\infty$, depending on the formulation, the original linear theory of Maxwell is restored.  

\noindent We also know that singularities are abundant in solutions in Einstein's General Relativity (GR). These can be identified as causing a divergence of curvature scalars or incompleteness of causal geodesics in certain regions of spacetime. The first black hole solution, the Schwarzschild solution, shows a divergence or breakdown of geodesics when the radial coordinate becomes zero. Further, the Hawking-Penrose singularity theorems \cite{Hawking1, Penrose, Hawking2} state that there are incomplete geodesics, provided certain energy conditions hold and additional constraints on the causal structure of spacetime are fulfilled  \cite{Senovilla}. Thus, spacetime singularities in Einstein's GR are not merely coordinate artefacts or coincidental results of extremely symmetrical solutions. Instead, they represent the limitations of classical GR. It is essential to tackle this problem in order to comprehend the ultimate result of gravitational collapse of initially stable structures.

\noindent There is widespread anticipation that quantum extensions of classical theories would remove or resolve the singularities present in those theories \cite{Crowther}. For instance, the divergence problem of Maxwell's electrodynamics can be cured by invoking quantum electrodynamics \cite{Lifshitz}. There are some effective models at various levels of string/M-theory too \cite{Sorokin, Fradkin}. Similar to linear electrodynamics, the singularity issue in classical GR may be approached in different ways. One possible direction is to invoke `quantum gravity', the quantum theory of classical GR, which is still in a state of `progress'. Second, one can modify Einstein's GR (by deforming the Einstein-Hilbert action) and look for a divergence-free solution in that specific theory of gravity. An alternative method involves exploring various nonlinear generalisations of Einstein-Maxwell theory. Thus, it is expected that the problem of spacetime singularities may be resolved by minimal coupling of a suitable NLE Lagrangian with gravity \cite{Ayon1, Ayon2}. Our work primarily revolves around this line of thought.

\noindent First, in 1968, Bardeen came up with his \emph{ad hoc} proposal of a singularity-free black hole solution \cite{Bardeen}. Subsequently, several suggestions of regular models have been reported in scientific publications \cite{Hayward, Roman, Dymnikova1, Dymnikova2, Bronnikov3, Carballo2, Bambi, Carballo3, Frolov3, Balart1, Frolov}. All of these black holes have been often referred to as `regular black holes'. As indicated earlier, it is found that one of the ways to understand regular solutions is the minimal coupling of Einstein's GR with a chosen version of the matter NLE Lagrangian \cite{Ayon3, Ayon4, Ayon5, Fan1, Bronnikov4, Bronnikov5, Smolic}. In recent times, there have been several new constructions of regular black holes with unique features \cite{Kolar, Bueno, Bokulic, Bonanno, Lessa, Estrada, Karakasis, Ovalle, Ara}.

\noindent The derivation of matter NLE Lagrangians usually involves reverse engineering of already proposed regular models as is done for a generic class of regular black holes in \cite{Fan, Toshmatov}, leaving them open to criticism from various angles. First, the dynamics of the NLE Lagrangians need to be clearly understood, at least in flat spacetime. The nonlinear evolution of waves, their scattering, birefringence, and other related effects are still not explored. Second, the matter NLE models contain a fractional power of the Maxwell invariant ($F=\frac{1}{4}F_{\mu\nu}F^{\mu\nu}$), which limits the consideration to only positive values of $F$. As a result, the electric field of a point charge ($F$ negative) cannot be described in the context of these Lagrangians. However, it is not an issue in constructing a regular black hole because a nonlinear magnetic monopole (where $F$ is positive) can sustain it. But, the minimal coupling of these Lagrangians with GR in the presence of an electric point charge 
remains out of scope. As electric charge exists in nature, such NLE models seem to have limitations. Moreover, the original motivation of nonlinear electrodynamics (elimination of the singularity in the Maxwell field -- as mentioned earlier) is not fulfilled by these NLE Lagrangians. Lastly, we also note that there are examples of electrically charged regular black holes \cite{Bronnikov1, Bronnikov2, Marco}, 
though the matter Lagrangian violates the Maxwellian limit, and the Lagrangian description is somewhat ambiguous (multi-valued functions).

\noindent Considering the issues mentioned above, we are motivated to construct a novel NLE Lagrangian model which follows the Maxwellian limit and is free from any fractional power of $F$. In other words, we aim to examine its coupling with gravity in order to create black hole solutions with both electric and magnetic characteristics. Before introducing a curved spacetime background, we analyse our proposed model in flat spacetime to establish its credibility as an alternative to the BI model. There is already a significant amount of research on this subject, specifically discussing the role of NLE models in sourcing various types of spacetime solutions \cite{Krug, Dey, Diaz, Aiello, Paul, Junior, Habibina}.

\noindent As discussed before, the primary objective of NLE was to eliminate the self-energy singularity in linear electrodynamics. However, we must consider to what extent we can benefit from this process and whether eliminating spacetime singularities is eventually possible. We argue that the NLE model presented in this paper could cure the singularities in Maxwell's electrodynamics in flat spacetime and also generate regular spacetimes when coupled to Einstein's GR in the presence of a suitable source charge. 
 
\noindent This article is organised as follows. In Section \ref{II}, we introduce the model and study its field equations in flat spacetime. In Section \ref{III}, the energy of a single point charge is calculated. Section \ref{IV} demonstrates the phenomenon of vacuum birefringence. In Section \ref{V}, we have studied the causality and unitarity conditions on the Lagrangian. The coupling of the Lagrangian with Einstein's GR is analysed in section \ref{VI}. Finally, we conclude in Section \ref{VII}.

\noindent We adopt the mostly positive metric signature $(-,+,+,+)$ and work in geometrical units, i.e. $G=c=1$. 

\section{Nonlinear electrodynamics model and field equations}\label{II}
\noindent This section introduces our general nonlinear Lagrangian and imposes restrictions on the model parameters to ensure its viability. In this work, we choose some feasible values for model parameters to continue our further analysis. We examine the nonlinear Maxwell-like equations in flat spacetime for the specific Lagrangian (having particular values of model parameters), focusing on their electrostatic limit. The variation of the electric field with radial distance, due to a point charge, is demonstrated analytically via a power series (in the weak field limit), as well as graphically.

\begin{figure}[h]
\centering
\includegraphics[width=0.6\textwidth]{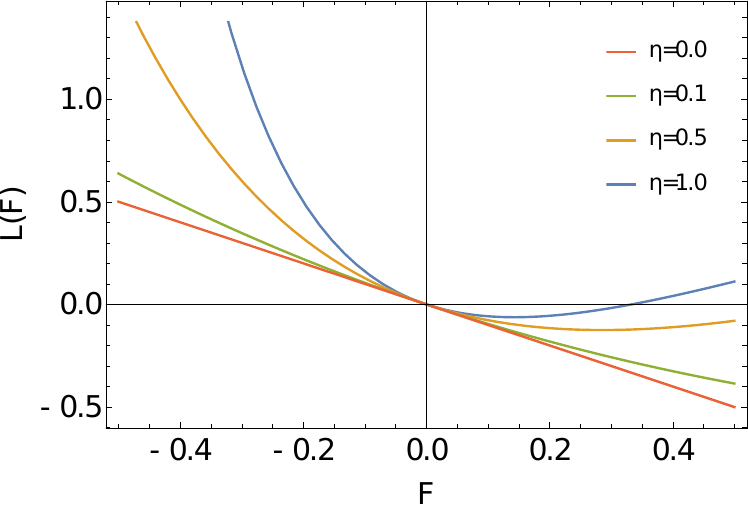}
\caption{Plot of the new NLE model Lagrangian $L(F)$ in Eq.(\ref{eq4}) for different values of $\eta$ along with the linear model ($\eta=0$) as a function of $F$, assuming $\gamma=1$}
\label{fig:lagrangian}
\end{figure}

\noindent Let us begin with Maxwell's Lagrangian, which is directly proportional to $F=\frac{1}{2}(\textbf{B}^2-\textbf{E}^2)$, i.e.,
\begin{equation}\label{eq1}
    L_{Maxwell}=-F
\end{equation}
As mentioned earlier, the BI model is defined by the following Lagrangian \cite{Born1, Born2}.
\begin{equation}\label{eq2}
    L_{BI}=\frac{2}{\beta}(1-\sqrt{1+\beta F})
\end{equation}
where $\beta$ is an arbitrary constant. In the weak field limit $(\beta F<<1)$, the Lagrangian in Eq.(\ref{eq2}) becomes,
\begin{equation}\label{eq3}
    L_{BI}\approx -F+\frac{1}{4}\beta F^2-\frac{1}{8}\beta^2 F^3+O(\beta^3 F^4)
\end{equation}
Therefore, it is evident that as $\beta\to 0$, we have the linear Maxwell Lagrangian. Let us now consider a more general Lagrangian, which defines our novel nonlinear electrodynamics model,
\begin{equation}\label{n1}
    L_{general}(F)=-\frac{(a F+1)^m}{\delta(b F+1)^n}(\beta F)^p
\end{equation}
where $m$, $n$ and $p$ are dimensionless constants, and $a$, $b$, $\beta$, $\delta$ are arbitrary parameters with dimensions of squre length. Further, in the weak field limit, Eq.(\ref{n1}) has the following form,
\begin{equation}\label{n2}
    L_{general}(F)\approx -c\left( F^p+ c_{1} F^{p+1}+ c_{2} F^{p+2}+ O(c_{3} F^{p+3})\right)
\end{equation}
The expression above indicates that the Maxwellian limit can be obtained when $p=1$. Since a viable matter must follow the Maxwellian limit, we are bound to obey this restriction. It will be shown in Section \ref{V} that the background electric field due to a point charge holds causality if $n\geq (m+1)$, $a\leq 0$ and $b\geq 0$. We choose the extreme condition $n=(m+1)$, particularly $m=1$, to continue our analysis. Moreover, in this study, we pick $a=-3b$, which represents certain unique features when coupled to Einstein's GR, as discussed in Section \ref{VI}. Hence, our chosen two-parameter Lagrangian, which obeys the Maxwellian limit, takes the following form,
\begin{equation}\label{eq4}
    L(F)=\frac{\gamma\left(3\eta F-1\right)F}{\left(1+\eta F\right)^2}
\end{equation}
where $\gamma$ $(=\beta/\delta)$ and $\eta$ are arbitrary parameters. For $\eta F << 1$, i.e. for a weak field, the above Lagrangian becomes,
\begin{equation}\label{eq5}
    L(F)\approx -\gamma F+5\gamma\eta F^2-9\gamma \eta^2 F^3+\gamma O(\eta^3 F^4)
\end{equation}
As expected, the BI Lagrangian (\ref{eq2}) as well as the new NLE model (\ref{eq4}) ($\gamma=1$) are schematically identical in the weak field limit (in the two series expansion (\ref{eq3}), (\ref{eq5}), all powers of $F$ are equal.). Additionally, by construction, both reduce to the standard Maxwellian theory (\ref{eq1}) when the nonlinear parameter vanishes. In Figure \ref{fig:lagrangian}, we provide a graphical analysis that compares the Lagrangian in Eq.(\ref{eq4}) with the linear model. It is apparent that for small values of $\eta$, the new NLE model exhibits nearly linear behaviour, replicating the limit of the BI model. However, the deviation from linearity becomes significant for higher values of $\eta$.

\noindent Next, the Euler-Lagrange equation of motion of the Lagrangian in Eq.(\ref{eq4}), in flat spacetime, can be written as,
\begin{equation}\label{eq6}
    \partial_{\mu}(L_{F}F^{\mu\nu})=0
\end{equation}
where,
\begin{equation}\label{eq7}
    L_{F}=\frac{\partial L}{\partial F}=\frac{\gamma(-1+7\eta F)}{\left(1+\eta F\right)^3}
\end{equation}
and $F_{\mu\nu}$ is the Maxwell field strength tensor. Now, by using the relation for the electric displacement field  $\textbf{D}=\partial L/\partial\textbf{E}$, one has, from Eq.(\ref{eq4}),
\begin{equation}\label{eq8}
    \textbf{D}=\gamma\frac{1-7\eta F}{(1+\eta F)^3}\textbf{E}
\end{equation}
where $\textbf{E}$ is the electric field. Eq.(\ref{eq8}) can be expressed in tensor form as $\textbf{D}_{i}=\epsilon_{i}{}^{j}\textbf{E}_{j}$, where $\epsilon_{ij}$ represents the electric permittivity tensor given below,
\begin{equation}\label{eq9}
    \epsilon_{ij}=\gamma\frac{1-7\eta F}{\left(1+\eta F\right)^3}\delta_{ij}
\end{equation}
Similarly, the magnetic field $\textbf{H}=-\partial L/\partial\textbf{B}$ for the same Lagrangian (\ref{eq4}) can be expressed in the following form,
\begin{equation}\label{eq10}
    \textbf{H}=\gamma\frac{1-7\eta F}{(1+\eta F)^3}\textbf{B}
\end{equation}
Now the magnetic induction field is $\textbf{B}_{i}=\mu_{i}{}^{j}\textbf{H}_{j}$, therefore the inverse of magnetic permeability tensor $(\mu^{-1})_{ij}$ becomes,
\begin{equation}\label{eq11}
    (\mu^{-1})_{ij}=\gamma\frac{1-7\eta F}{\left(1+\eta F\right)^3}\delta_{ij}
\end{equation}
Further, the Euler-Lagrange equation of motion (\ref{eq6}) can be reformulated as the first set of source-free Maxwell's equations using Eqs. (\ref{eq8}) and (\ref{eq10}),
\begin{equation}\label{eq12}
    \nabla\cdot\textbf{D}=0, \hspace{20pt} \frac{\partial\textbf{D}}{\partial t}-\nabla\times\textbf{H}=0
\end{equation}
And, from the Bianchi identity $\partial_{\mu}\Tilde{F}^{\mu\nu}=0$, we have the other set,
\begin{equation}\label{eq13}
    \nabla\cdot\textbf{B}=0,\hspace{20pt} \frac{\partial \textbf{B}}{\partial t}+\nabla\times\textbf{E}=0
\end{equation}
where $\Tilde{F}^{\mu\nu}$ is the dual of the field strength tensor. It is apparent that the Maxwell's equations (\ref{eq12}) and (\ref{eq13}) incorporate the electric permittivity tensor $\epsilon_{ij}$ and the magnetic permeability tensor $\mu_{ij}$, both of which depend on the Maxwell invariant $F$. This means that the nonlinearity of the Lagrangian (\ref{eq4}) is encoded via an `anisotropic' medium with specific properties, as evident from Eqs. (\ref{eq9}) and (\ref{eq11}). It is important to note that $\textbf{D}\cdot\textbf{H}\neq\textbf{E}\cdot\textbf{B}$, indicating a violation of duality symmetry \cite{Gibbons}. Setting the values of $\eta$ and $\gamma$ to 0 and 1, respectively, leads to the recovery of Maxwell's electrodynamics and restoration of duality symmetry. It is worth mentioning that the BI model maintains duality symmetry, but in Quantum Electrodynamics (QED), this symmetry is broken by one-loop quantum corrections \cite{Kruglov3}.

\vspace{0.3cm}
\noindent\textbf{Electrostatic limit:}
Let us examine the field equations of electrostatics when both the magnetic induction ($\textbf{B}$) and the magnetic field intensity ($\textbf{H}$) are equal to zero. The equation for a point charge can be expressed as,
\begin{equation}\label{eq14}
    \nabla\cdot\textbf{D}=e\delta(r)
\end{equation}
where $e$ is the electric charge. The solution of the above equation can be written as,
\begin{equation}\label{eq15}
    \textbf{D}=\frac{e}{4\pi r^3}\textbf{r}
\end{equation}
With the help of Eq.(\ref{eq8}), the above equation becomes,
\begin{equation}\label{eq16}
    E+\frac{7}{2}\eta E^{3}=\frac{e}{4\gamma\pi r^2}\left(1-\frac{\eta}{2} E^{2}\right)^3
\end{equation}
where we have substituted $F=-E^{2}/2$ and $E$ is the magnitude of electric field $\textbf{E}$. As the radial coordinate $r\in \mathbb{R}^{+}$, the above equation indicates a restriction on the domain of $F$, i.e. $F$ is larger than $-\frac{1}{\eta}$. However, there is no such restriction on $F$ for the purely magnetic case. Therefore, in general, $F$ can vary between $-\frac{1}{\eta}$ to $\infty$.  One may note that in the Maxwellian limit, the boundary of $F$ extends to $(-\infty,+\infty)$ as $\eta \to 0$, which is expected.

\begin{figure}[h]
\centering
\includegraphics[width=0.6\textwidth]{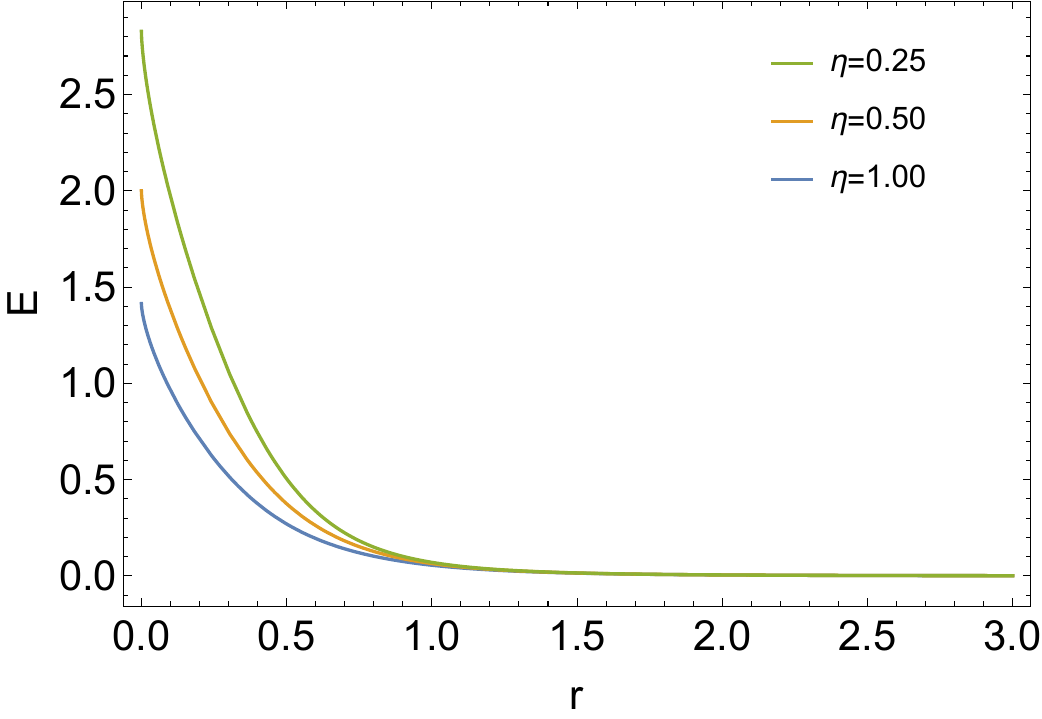}
\caption{Plot of the electric field of a point charge as a function of $r$ for different values of $\eta$, assuming $\gamma=1$ and $e=1$}
\label{fig:electric}
\end{figure}

\noindent Let us try to obtain an analytical solution of $E(r)$ from the above polynomial equation (\ref{eq16}). This is challenging for an arbitrary value of $\eta$. However, in the weak field limit $(\eta F<<1)$, one can express $E(r)$ in a perturbative series in $\eta$. In this limit, the total electric field $E(r)$ can be written as,
\begin{equation}\label{eq17}
    E=E_{(0)}+\eta E_{(1)}+\eta^{2}E_{(2)}+ O(\eta^{3})
\end{equation}
where $E_{(1)}$ and $E_{(2)}$ represent the first and second-order corrections to the electric field $E_{(0)}$, respectively. Although there are higher-order corrections, their contribution to the total electric field is negligible. Here, $E_{(0)}$, $E_{(1)}$, and $E_{(2)}$ have the units of length inverse, the cube of length inverse, and the fifth power of length inverse, respectively. One can substitute the above expansion of $E(r)$ into Eq.(\ref{eq16}). And, by comparing the resultant expressions at different orders of $\eta$, we have,

\noindent Zeroth-order,
\begin{equation}\label{eq18}
    E_{(0)}=\frac{e}{4\pi\gamma r^{2}}
\end{equation}
First-order,
\begin{equation}\label{eq19}
    E_{(1)}=-\frac{7}{2}E_{(0)}^{3}-\frac{e}{4\pi\gamma r^{2}}\frac{3}{2}E_{(0)}^{2}
\end{equation}
Second-order,
\begin{equation}\label{eq20}
    E_{(2)}=-\frac{21}{2}E_{(0)}^{2}E_{(1)}+\frac{e}{4\pi\gamma r^2}\left(-3 E_{(0)}E_{(1)}+\frac{3}{4}E_{(0)}^{4}\right)
\end{equation}
Solving the above three equations, one can obtain the correction terms to the electric field $E_{(0)}$. As a result, we are led to the following expression for the total electric field of a point charge (in the weak field limit),
\begin{equation}\label{eq22}
    E\approx \frac{e}{4\pi\gamma r^{2}}-5\eta\left(\frac{e}{4\pi\gamma r^{2}}\right)^{3}+\frac{273}{4}\eta^{2}\left(\frac{e}{4\pi\gamma r^{2}}\right)^{5}+ O(\eta^{3})
\end{equation}
One may note that the zeroth order contribution is identical to that for Maxwell's electrodynamics. Although from Eq.(\ref{eq22}), the finiteness of the electric field is still unclear, one can get an analytical understanding of the electric field caused by a point charge. However, Eq.(\ref{eq16}) indicates the following maximum value of electric field $E(r)$ for small values of $r$ and any arbitrary value of $\eta$,
\begin{equation}\label{eq23}
    E_{max}=\sqrt{\frac{2}{\eta}}
\end{equation}
Therefore, the maximum electric field of a point charge has a finite value in our model, similar to BI electrodynamics. Unlike linear electrodynamics, whose electric field strength exhibits a singularity leading to an unlimited amount of self-energy associated with a point charge, our new NLE model eliminates the Maxwellian singularity. In Figure \ref{fig:electric}, we present the variation of the electric field by plotting Eq.(\ref{eq16}) with radial distance for different values of $\eta$ (not necessarily in the weak field) to demonstrate its finiteness. One may note that for small values of $\eta$, the maximum value of the electric field increases, and as $\eta\to 0$, it will eventually diverge.

\section{Energy of a point charge}\label{III}
\begin{figure}[h]
\centering
\includegraphics[width=0.6\textwidth]{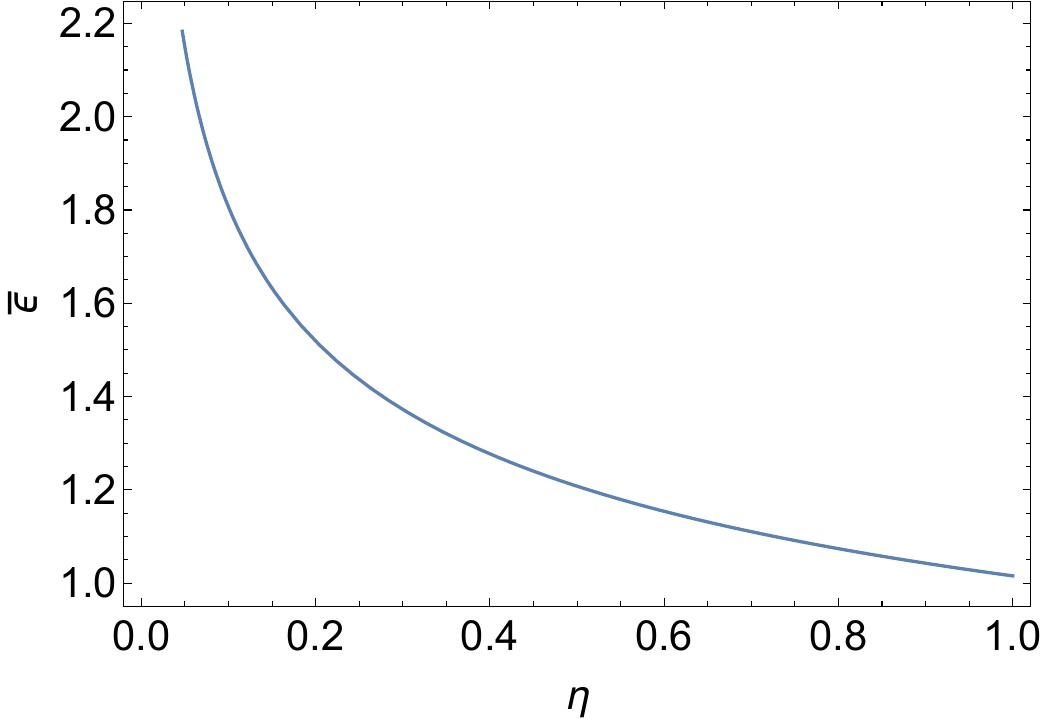}
\caption{Variation of electrostatic energy of a point charge ($\Bar{\mathbf{\epsilon}}$) with $\eta$, reflects its finiteness.}
\label{fig:energy}
\end{figure}

\noindent We now turn to the total electric energy of the point charge. It is known that energy density in a classical field theory can be obtained by taking the negative of the $T^{t}_{t}$ component of the energy-momentum tensor $T_{\mu}^{\nu}$. In the literature, there are several inequivalent definitions of stress-energy tensors in Minkowski spacetime, such as the canonical electromagnetic stress-energy tensor \cite{Blaschke}, the Belinfante–Rosenfeld stress-energy tensor \cite{Belinfante}, and the Hilbert stress-energy tensor \cite{Blaschke}. We consider the Hilbert stress-energy tensor in flat spacetime, which is both symmetric and gauge invariant. It is defined as,
\begin{equation}\label{eq24}
    T_{\mu\nu}^{H}=-\frac{2}{\sqrt{-g}}\left.\left(\frac{\partial\left(\sqrt{-g}L(F)\right)}{\partial g^{\mu\nu}}\right)\right|_{g=\eta}
\end{equation}
where we have derived the stress-energy tensor in a general `curved spacetime' defined by the metric tensor $g_{\mu\nu}$. Thus, the Hilbert stress-energy tensor corresponding to the Lagrangian in Eq.(\ref{eq4}) can be written as,
\begin{equation}\label{eq25}
    T_{\mu\nu}^{H}=\eta_{\mu\nu}L(F)-L_{F}F_{\mu}^{\alpha}F_{\nu\alpha}
\end{equation}
where $L_{F}$ is defined in Eq.(\ref{eq7}). Consequently, the electric energy density becomes,
\begin{equation}\label{eq26}
    \rho=-T_{t}^{t}=-L_{F}E^{2}-L(F)=\frac{\gamma E^{2}\left(1+\frac{3}{2}\eta E^{2}\left(4+\frac{\eta}{2}E^{2}\right)\right)}{2\left(1-\frac{\eta}{2}E^{2}\right)^{3}}
\end{equation}
The total electric energy is given by
\begin{equation}\label{eq27}
    \mathbf{\epsilon}=4\pi\int_{0}^{\infty}\rho(r) r^{2} dr
\end{equation}
Using Eq.(\ref{eq16}), it is possible to transform the above integral over `$r$' into an integral over the variable `$E$'. Following this transformation, the above integral becomes,
\begin{equation}\label{eq28}
    \mathbf{\epsilon}=\frac{e^{3/2}}{\sqrt{4\pi\gamma}}\int_{0}^{\sqrt{\frac{2}{\eta}}}\frac{\sqrt{(2-\eta E^{2})}\left(4+3\eta E^{2}(8+\eta E^{2})\right)\left(4+\eta E^{2}(52+21\eta E^{2})\right)}{16\sqrt{E}(2+7\eta E^{2})^{5/2}} dE
\end{equation}
One can numerically solve the above integral for different values of $\eta$. This gives us the self-energy of a point charge, which depends on the electric charge $e$ as well as the model parameters $\gamma$ and $\eta$. In Figure \ref{fig:energy}, the variation of the new variable $\Bar{\mathbf{\epsilon}}\equiv \mathbf{\epsilon}\sqrt{4\pi\gamma}/e^{3/2}$, which is the scaled version of total energy $\epsilon$, with $\eta$ is presented. It is evident that for an arbitrary value of $\eta$, the energy of a point charge is finite. And in the Maxwell limit ($\eta\to 0$), the integral in Eq.(\ref{eq28}) becomes,
\begin{equation}
    \mathbf{\epsilon}\approx\frac{e^{3/2}}{\sqrt{4\pi\gamma}}\left[\frac{\sqrt{E}}{2}\right]_{0}^{\infty}
\end{equation}
which shows an infinite amount of energy (as expected). Thus, the electrostatic energy of a point charge is finite in our model ($\eta\neq 0$), like it is in the well-known BI model.

\section{Vacuum birefringence}\label{IV}
\noindent In the realm of nonlinear electrodynamics (NLE), it is well recognised that photons do not travel along null geodesics of the background Minkowski geometry. Instead, they follow a different trajectory in an `effective geometry' \cite{Novello2, Novello}. The effective metric depends on the dynamics of the background electromagnetic field. It is possible to describe the null geodesics in the effective geometry as wave propagation, governed by Maxwellian electrodynamics, in a nonlinear dielectric medium \cite{Novello}. Therefore, the nonlinearity in the Lagrangian has a \textit{medium interpretation}, mentioned previously. Light may be considered as electromagnetic waves moving through a classical dispersive medium. Hence, several phenomena usually associated with a dielectric medium will occur even in vacuum in NLE.

\noindent The phenomenon of vacuum birefringence in quantum electrodynamics (QED) was examined in references \cite{Adler, Biswas}. Recently, several studies have been conducted on causal structure, birefringence and light propagation in a general nonlinear electrodynamics Lagrangian \cite{Tomizawa, Schellstede, Melo, Russo, Russo2}. Further, vacuum birefringence in BI electrodynamics has been discussed in \cite{Kruglov4, Kruglov5}. Let us now investigate this effect in our NLE model. We consider the propagation of plane electromagnetic waves $(\textbf{e},\textbf{b})$ travelling in the $z$ direction between two plates of a capacitor having a constant and uniform electric field along the $x$ direction, i.e. the external field $\Bar{\textbf{E}}=(\Bar{E},0,0)$. The total electromagnetic field is the sum of the light field $(\textbf{e},\textbf{b})$ and the background electric field, i.e.  $\textbf{E}=\textbf{e}+\Bar{\textbf{E}}$, $\textbf{B}=\textbf{b}$.  We assume that the light fields are much weaker than the background electric field. Thus, the Lagrangian in Eq.(\ref{eq4}) becomes:
\begin{equation}\label{eq29}
    L(\textbf{e}+\Bar{\textbf{E}},\textbf{b})=\gamma\frac{\Big\lbrace\frac{3}{2}\eta(\textbf{b}^{2}-(\textbf{e}+\Bar{\textbf{E}})^{2})-1\Big\rbrace(\textbf{b}^{2}-(\textbf{e}+\Bar{\textbf{E}})^{2})}{2\big\lbrace 1+\frac{\eta}{2}(\textbf{b}^{2}-(\textbf{e}+\Bar{\textbf{E}})^{2})\big\rbrace^{2}}
\end{equation}
However, neglecting the higher orders of the light field in the above Lagrangian, we are left only with quadratic terms in $\textbf{e}$ and $\textbf{b}$, which yields,
\begin{equation}\label{eq30}
    L^{(2)}(\textbf{e}+\Bar{\textbf{E}},\textbf{b})=\frac{\gamma\eta\left(5+\frac{7}{2}\eta\Bar{\textbf{E}}^{2}\right)}{\left(1-\frac{\eta}{2}\Bar{\textbf{E}}^{2}\right)^{4}} (\textbf{e}.\Bar{\textbf{E}})^{2}-\frac{1}{2}
   \gamma\frac{\left(1+\frac{7}{2}\eta\Bar{\textbf{E}}^{2}\right)}{\left(1-\frac{\eta}{2}\Bar{\textbf{E}}^{2}\right)^{3}} (\textbf{b}^{2}-\textbf{e}^{2})
\end{equation}
Next, it is possible to find the linearised electric displacement and magnetic field vectors on the background electric field from the above Lagrangian (\ref{eq30}),
\begin{equation}\label{eq31}
       d_{i}=\frac{\partial L^{(2)}}{\partial e^{i}}=(\alpha\delta_{i}^{j}+\beta \Bar{E}_{i}\Bar{E}^{j})e_{j}, \hspace{0.5cm} h_{i}=-\frac{\partial L^{(2)}}{\partial b_{i}}=\alpha\delta_{i}^{j}b_{j}
\end{equation}
where, 
\begin{equation}\label{eq32}
    \beta=\frac{2\gamma\eta\left(5+\frac{7}{2}\eta\Bar{\textbf{E}}^{2}\right)}{\left(1-\frac{\eta}{2}\Bar{\textbf{E}}^{2}\right)^{4}}, \hspace{1cm} \alpha=
   \gamma\frac{\left(1+\frac{7}{2}\eta\Bar{\textbf{E}}^{2}\right)}{\left(1-\frac{\eta}{2}\Bar{\textbf{E}}^{2}\right)^{3}}
\end{equation}
Therefore, only linear terms in $\textbf{e}$ and $\textbf{b}$ are left in Eq.(\ref{eq31}). Using the relations $d_{i}=\epsilon^{j}_{i}e_{j}$, $h_{i}=(\mu^{-1})_{i}^{j}b_{j}$, electric permittivity and permeability tensors can be written as,
\begin{equation}\label{eq33}
        \epsilon_{ij}=\alpha\delta_{ij}+\beta \Bar{E}_{i}\Bar{E}_{j},\hspace{0.5cm}(\mu^{-1})_{ij}=\alpha\delta_{ij}
\end{equation}
The Maxwell equations obeyed by a plane electromagnetic wave imply,
\begin{equation}\label{eq34}   k_{i}d^{i}=k_{i}b^{i}=0,\hspace{1cm}\textbf{k}\times\textbf{e}=\omega\textbf{b},\hspace{1cm}\textbf{k}\times\textbf{h}=-\omega\textbf{d}
\end{equation}
where $\textbf{k}$ is the wave vector. Substituting the vectors $d_{i}=\epsilon_{i}^{j}e_{j}$ and $h_{i}=(\mu^{-1})_{i}^{j}b_{j}$ in Eq.(\ref{eq34}), one obtains the wave equation for the electric field \textbf{e} as follows \cite{Kruglov5},
\begin{equation}\label{eq35}
    \Big\{\boldsymbol{\epsilon}^{ijk}\boldsymbol{\epsilon}_{lmn}\left(\mu^{-1}\right)^{l}_{k}k_{j}k^{m}+\omega^{2}\epsilon^{i}_{n}\Big\}e^{n}=0
\end{equation}
where $\boldsymbol{\epsilon}_{ijk}$ is the antisymmetric tensor. The above equation can be converted into the matrix equation,
\begin{equation}\label{eq36}
    \Lambda \textbf{e}=0
\end{equation}
where,
\begin{gather*}\label{eq37}
 \Lambda
 =
  \begin{bmatrix}
   -k^{2}\alpha+\omega^{2}(\alpha+\beta\Bar{E}^{2}) & 0 & 0 \\
   0 & -k^{2}\alpha+\omega^{2}\alpha & 0 \\
   0 & 0 & \omega^{2}\alpha
   \end{bmatrix}
\end{gather*}
Thus, when the matrix determinant is zero, there are nontrivial solutions to the homogeneous matrix equation (\ref{eq36}). As a result, we have two modes corresponding to the electric field (polarisation direction) of the plane wave. These modes define the dispersion relations, which may be expressed in terms of indices of refraction in the following way,
\begin{equation}\label{eq38}
        n_{||}=\sqrt{1+\frac{\beta}{\alpha}\Bar{E}^{2}}, \hspace{1cm} n_{\perp}=1
\end{equation}
where the index of refraction is defined, as usual, by $n=k/\omega$. Hence, electromagnetic waves with different polarisations have different velocities: $v_{||}=n_{||}^{-1}$ and $v_{\perp}=1$. Here, parallel and perpendicular polarisations are defined with respect to the background uniform electric field. It is essential to observe that $\alpha$ and $\beta$ are consistently positive. Consequently, the parallel phase velocity is less than $1$, while the perpendicular velocity equals $1$. These results show the phenomenon of vacuum birefringence, which, in this case, is a consequence of the nonlinear nature of our model. As $\eta$ approaches 0, $\beta$ vanishes, and the phenomenon of vacuum birefringence disappears, i.e., two modes have the same velocity equal to one.

\section{Causality and unitarity conditions on the Lagrangian }\label{V}
\noindent Let us now examine the causality and unitarity restrictions on the Lagrangian itself as obtained by Shabad and Usov \cite{Shabad}. The NLE models are considered valid when the requirements of causality and unitarity hold. If the Lagrangian satisfies the following inequalities:
\begin{equation}\label{eq39}
    L_{F}\leq 0, \hspace{1cm} L_{FF}\geq 0,\hspace{1cm} L_{F}+2FL_{FF}\leq 0
\end{equation}
group velocity of elementary excitations over a background field is not greater than the speed of light in vacuum. Here, subscripts refer to the partial derivatives of the Lagrangian with respect to the invariant $F$. Further, if Eq.(\ref{eq39}) holds, the particles always possess a positive amount of kinetic energy or at least a minimum positive threshold for such energy. 

\noindent We analyse the causality and unitarity conditions for our general Lagrangian mentioned earlier in Eq.(\ref{n1}). We explain how these restrictions guide us in selecting the model parameters at the beginning in Section \ref{II}.  Subsequently, we demonstrate causality and unitarity conditions for the Lagrangian in Eq.(\ref{eq4}) as a special case of the general one in Eq.(\ref{n1}). The derivatives of the general Lagrangian in Eq.(\ref{n1}) are as follows,

\begin{eqnarray}\label{n3}
    L_{F}&=&\gamma\frac{abF^{2}(n-m-1)+F\lbrace b(n-1)-a(m+1)\rbrace-1}{(1+aF)^{1-m}(1+bF)^{1+n}} \nonumber \\
    && \nonumber \\
    L_{FF}&=&\frac{\gamma}{(1+aF)^{2-m}(1+bF)^{2+n}}\Bigg\lbrace -ma(1+bF)^{2}(2+aF(1+m))\nonumber \\
    && \nonumber \\
    && 
    +nb(1+aF)(2+bF+aF(2+bF+2m+2mbF))-b^{2}F(1+aF)^{2}n^{2}\Bigg\rbrace
\end{eqnarray}
where we assumed that $m$ and $n$ are greater than or equal to zero.

\subsection{Causality and Unitarity for the Electric part }\label{sub1}
\noindent Now, let us look above inequalities (\ref{eq39}) by considering only the electric field, i.e. $B=0$. It is evident from the derivatives of the general Lagrangian, as shown in Eq.(\ref{n3}), the first two inequalities of Eq.(\ref{eq39}) hold for,
\begin{equation}\label{n4}
    n\geq m+1; \hspace{2cm} a\leq 0, b\geq 0
\end{equation}
Here $F$ is a negative quantity ($B=0$), which implies an additional restriction that $F>-\frac{1}{b}$ (mentioned earlier in Section \ref{II}, range of $F$). Note that the above bounds on model parameters in Eq.(\ref{n4}) are sufficient to satisfy the third inequality in Eq.(\ref{eq39}).

\subsection{Causality and Unitarity for the Magnetic case }\label{sub2}
\noindent In a similar manner, considering only magnetic field and analysing the first condition of Eq.(\ref{eq39}), we obtain
\begin{equation}\label{n5}
    n\geq m+1; \hspace{2cm} a\geq 0, b\leq 0
\end{equation}
where $F$ is positive ($E=0$). However, we found that the second inequality of Eq.(\ref{eq39}) disregards the above restriction (\ref{n5}) on model parameters as obtained from the first one. Hence, there is no common range of model parameters that satisfies the first two inequalities of Eq.(\ref{eq39}). Moreover, from the third condition, it is not possible to draw any conclusion as it depends on the first two. Hence, the restriction in Eq.(\ref{n5}) is not a well-behaved one.

\noindent Therefore, as stated at the very beginning of the article, we opted to work with the bounds mentioned in Eq.(\ref{n4}), in particular $m=1$ and $a=-3b$, resulting in our Lagrangian model (\ref{eq4}). Below, we have examined the causality and unitarity conditions for our specific Lagrangian model (\ref{eq4}).

\subsection{Causality and Unitarity for $n=m+1$ and $a=-3b$}\label{sub3}
\noindent For these particular parameter values, the derivatives of Lagrangian (\ref{eq4}) are,
\begin{equation}\label{eq40}
    L_{F}=\frac{\gamma(-1+7\eta F)}{(1+\eta F)^{3}}, \hspace{1cm} L_{FF}=\frac{2\gamma\eta(5-7\eta F)}{(1+\eta F)^{4}}
\end{equation}
The third inequality in (\ref{eq39}) is,
\begin{equation}\label{eq41}
    L_{F}+2FL_{FF}=\gamma\frac{-1+\eta F(26-21\eta F)}{(1+\eta F)^{4}}
\end{equation}
For only electric field ($B=0$), causality and unitarity conditions are,
\begin{equation}\label{eq42}
    -\frac{2+7\eta E^{2}}{(2-\eta E^{2})^{3}}\leq 0, \hspace{0.7cm} \frac{10+7\eta E^{2}}{(2-\eta E^{2})^{4}}\geq 0, \hspace{0.7cm} \frac{-4-\eta E^{2}(52+21\eta E^{2})}{(2-\eta E^{2})^{4}}\leq 0
\end{equation}
all of these conditions states that $E<\sqrt{\frac{2}{\eta}}$. From equation (\ref{eq23}), we have the maximum value of the electric field as $E_{max}=\sqrt{\frac{2}{\eta}}$. Thus, the above requirements for causality and unitarity (\ref{eq39}) are upheld for all permissible values of $E$. Therefore, we may conclude that our NLE model satisfies both the causality and unitarity constraints for all allowed background electric fields as expected from the general analysis in subsection \ref{sub1}. 

\noindent Similarly, for only magnetic fields, analysing the first two conditions of Eq.(\ref{eq39}), we obtain
\begin{equation}
    \left(-1+\frac{7}{2}\eta B^{2}\right)\leq 0, \hspace{1cm} \left(5-\frac{7}{2}\eta B^{2}\right)\geq 0
\end{equation}
These two inequalities indicate $F<\frac{1}{7\eta}$. Hence, for a restricted range of the magnetic field, which depends on the value of $\eta$, our model (\ref{eq4}) obeys causality. 

\noindent Further, the unitarity condition gives,
\begin{equation}
    -2+\eta B^{2}\left(26-\frac{21}{2}\eta B^{2}\right)\leq 0
\end{equation}
It suggest the following domain of $F$, i.e. $\frac{13-2\sqrt{37}}{21}<\eta F<\frac{13+2\sqrt{37}}{21}$. Thus, causality and unitarity conditions are obeyed over a restricted range for purely magnetic fields.

\noindent In the above four sections, we have analysed our new NLE model in flat spacetime. Our proposed model removes the divergence problem in linear electrodynamics and exhibits vacuum birefringence (when $\eta\neq 0$). The conditions on causality and unitarity also hold. Therefore, it is inferred that the proposed NLE model exhibits satisfactory behaviour and can be considered as a matter source in a gravity-coupled scenario.

\section{The new NLE coupled to general relativity}\label{VI}
\noindent Let us now minimally couple the Lagrangian $L(F)$ (\ref{eq4}) with gravity through the action,
\begin{equation}\label{eq46}
    I=\int d^{4}x\sqrt{-g}\left(\frac{R}{\kappa}+L(F)\right)
\end{equation}
and look for different spacetime solutions. Here, $R$ is the Ricci scalar. The covariant equations of motion (field variation and metric variation) which emerge from the above action are given as,
\begin{equation}
    \begin{aligned}
    \nabla_{\mu}\left(\frac{\partial L}{\partial F}F^{\mu \nu}\right)=0 \\
    R_{\mu \nu}-\frac{1}{2}g_{\mu \nu}R=\kappa T_{\mu \nu}
\end{aligned}
\end{equation}
where $R_{\mu\nu}$ is the Ricci tensor and $T_{\mu\nu}$ is the Hilbert energy-momentum tensor expressed in a curved spacetime as,
\begin{equation}\label{eq49}
    T_{\mu}^{\nu}=L\delta_{\mu}^{\nu}-L_{F}F_{\mu\lambda}F^{\nu\lambda}
\end{equation}
Let us choose a spherically symmetric, static spacetime with the line element
\begin{equation}\label{eq50}
    ds^{2}=-f(r) dt^{2}+\frac{1}{f(r)} dr^{2}+r^{2}\left(d\theta^{2}+\sin^{2}{\theta}d\phi^{2}\right)
\end{equation}
We also assume only $F_{tr}$ and $F_{\theta\phi}$ as nonzero components in $F_{\mu\nu}$. $F_{tr}=-F_{rt}$ represents a radial electric field and $F_{\theta\phi}=-F_{\phi\theta}$ is a radial magnetic field. Therefore, the nonzero components of the stress energy-momentum tensor in Eq.(\ref{eq49}) turn out to be
\begin{equation}\label{eq51}
   \begin{aligned}
        T_{t}^{t}=T_{r}^{r}=&L(F)-L_{F}F_{tr}F^{tr}\\
        T_{\theta}^{\theta}=T_{\phi}^{\phi}=&L(F)-L_{F}F_{\theta\phi}F^{\theta\phi}
   \end{aligned}
\end{equation}
Consequently, three distinct spacetime solutions are possible: (A) exclusively magnetic spacetime, (B) solely electric spacetime, and (C) dyonic spacetime.  In this work, we will focus only on the purely magnetic and purely electric solutions. The dyonic solution is left for future work.
\subsection{The magnetic regular black hole solution}\label{VIA}
\noindent A purely magnetic solution arises from a magnetic field only $(F_{tr}=0)$, given by the nonzero Maxwell tensor component $F_{\theta\phi} =-q_m sin{\theta}$. Here, $q_m$ is a constant, which may be understood as the total charge of a magnetic monopole that results in a radial magnetic field $B_r=\frac{q_m}{r^2}$ (unit of length inverse). Consequently, the Maxwell invariant $F$ becomes as $F=\frac{q_{m}^{2}}{2r^{4}}$. It is important to note that the matter field due to a magnetic monopole is singular at $r=0$. The energy-momentum tensor (\ref{eq51}) in the presence of such a magnetic monopole may be written as follows,
\begin{equation}\label{eq52}
    T_{t}^{t}=T_{r}^{r}=\frac{\gamma q_{m}^{2}(3\eta q_{m}^2-2r^{4})}{(2r^{4}+\eta q_{m}^{2})^{2}}
\end{equation}
and, 
\begin{equation}\label{eq53}
    T_{\theta}^{\theta}=T_{\phi}^{\phi}=\frac{\gamma q_{m}^{2}(4r^{8}-24\eta q_{m}^{2}r^{4}+3\eta^{2}q_{m}^{4})}{(2r^{4}+\eta q_{m}^{2})^{3}}
\end{equation}
On the other hand, Einstein tensor $G_{\mu}^{\nu}$ for the line element (\ref{eq50}) is obtained as
\begin{equation}\label{eq54}
    G_{\mu}^{\nu}=diag[\frac{f^{\prime}}{r}+\frac{f-1}{r^{2}},\frac{f^{\prime}}{r}+\frac{f-1}{r^{2}},\frac{f^{\prime\prime}}{2}+\frac{f^{\prime}}{r},\frac{f^{\prime\prime}}{2}+\frac{f^{\prime}}{r}]
\end{equation}
Here, $\lbrace\prime\rbrace$ denotes the radial derivative of the metric function $f(r)$. The `$tt$' or `$rr$' component of the Einstein-nonlinear-Maxwell equations reduce to
\begin{equation}\label{eq55}
    \frac{f^{\prime}}{r}+\frac{f-1}{r^{2}}=\kappa\frac{\gamma q_{m}^{2}(3\eta q_{m}^2-2r^{4})}{(2r^{4}+\eta q_{m}^{2})^{2}}
\end{equation}
By solving the above differential equation, we find that the metric function $f(r)$ is
\begin{equation}\label{eq56}
    f(r)=1+\frac{c_{0}}{r}+\frac{\kappa\gamma q_{m}^{2}r^{2}}{2r^{4}+\eta q_{m}^{2}}
\end{equation}
where $c_0$ is the integration constant. It can be shown that the above metric function solves the other components of the Einstein-nonlinear-Maxwell equation in a straightforward way.

\noindent One can choose the integration constant and model parameters, which may lead to spacetimes representing a family of regular black holes. The choices are, 
\begin{equation*}\label{eq57}
    c_{0}=0,\hspace{1cm}\gamma=-\frac{2 b_{0}^{2}}{\kappa q_{m}^{2}}, \hspace{1cm} \eta=\frac{2g^{4}}{q_{m}^{2}}
\end{equation*}
where $b_{0}$ and $g$ are constants of dimension of length. Therefore, the line element may be rewritten as,
\begin{equation}\label{eq58}
    ds^{2}=-\left(1-\frac{b_{0}^{2}r^{2}}{r^{4}+g^{4}}\right)dt^{2}+\left(1-\frac{b_{0}^{2}r^{2}}{r^{4}+g^{4}}\right)^{-1}dr^{2}+r^{2}\left(d\theta^{2}+\sin^{2}{\theta}d\phi^{2}\right)
\end{equation}
One may note that the above spacetime metric is asymptotically flat, i.e.
$$g_{tt}\to -1\hspace{1cm} \text{and} \hspace{1cm}g_{rr}\to 1\hspace{1cm} \text{as} \hspace{1cm}r\to \infty$$
At very small values of $r$, it behaves like de-Sitter space
$$g_{tt}\to -(1-c^{2}r^{2})\hspace{1cm}\text{and} \hspace{1cm}g^{rr}\to(1-c^{2}r^{2})\hspace{1cm} \text{as}\hspace{1cm}r\to 0$$
where $c$ is a constant. We claim that the metric in Eq.(\ref{eq58}) represents a family of regular spacetimes for certain parameter values. This may be confirmed by looking at how the curvature scalars behave with the radial coordinate $r$, as is shown later. Moreover, it should be noted that the metric mentioned above is different from all the well-known regular black hole spacetimes proposed by Bardeen, Hayward or others \cite{Bardeen, Hayward, Roman, Dymnikova1, Dymnikova2, Bronnikov3}. When the regularising parameter is eliminated, known regular black hole spacetimes reduce to the conventional Schwarzschild line element. Here, when $g=0$, the above metric in Eq.(\ref{eq58}) reduces to mutated, massless singular Reissner–Nordstr\"om (RN) solution with an `imaginary charge', which may be interpreted in different ways \cite{Rosen}. 

\noindent In order to understand the geometry in Eq.(\ref{eq58}) better, we first examine the roots of the equation $g_{tt}=0$. These roots correspond to null hyper-surfaces and indicate the position of the black hole horizon.
\begin{figure}[h]
\centering
\includegraphics[width=0.6\textwidth]{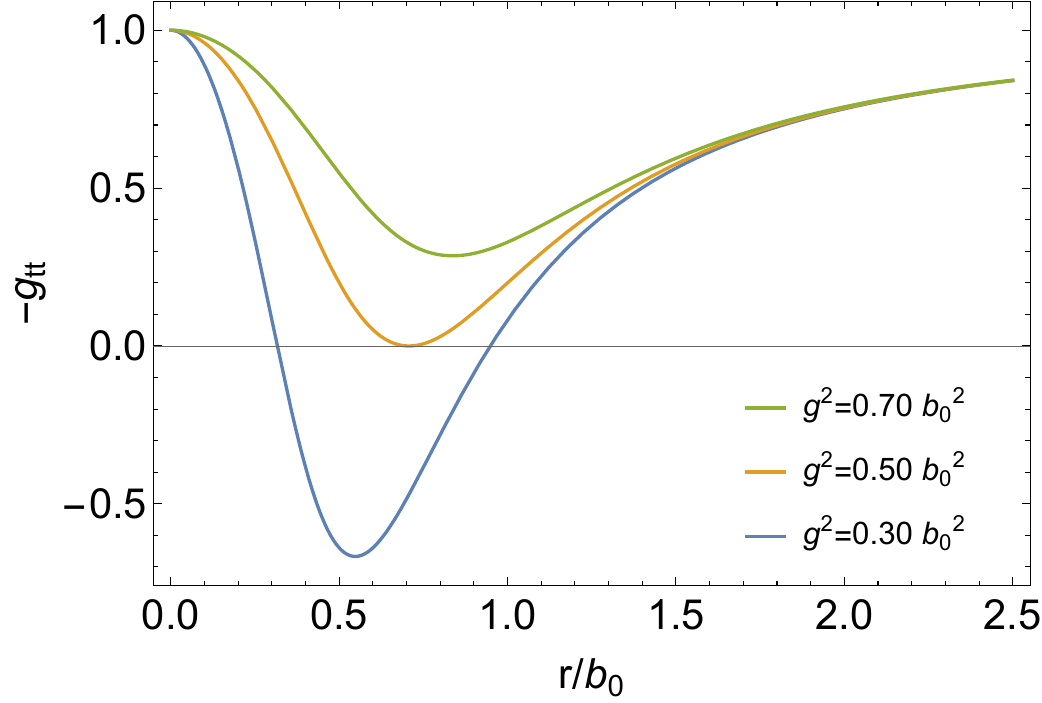}
\caption{Plot of redshift function with $r/b_{0}$ for different values of $g^{2}$}
\label{fig:horizon}
\end{figure}
Based on the horizon equation (explicit calculation not shown here) and with the help of Figure \ref{fig:horizon}, we may conclude that the above geometry reflects a family of black holes with a double horizon, provided that the metric parameters satisfy the condition $0<g^2<0.5b_0^2$. A single horizon black hole is obtained when $g^2=0.5 b_0^{2}$. Moreover, we have a spacetime without a horizon if $g^2>0.5b_0^2$. Notably, when $g^2=0$, the above spacetime corresponds to a black hole with a single horizon.

\subsubsection{\textbf{Regularity of curvature tensors and invariants}}
\noindent Are the above black holes or horizon-less spacetimes nonsingular? To examine the singular/regular characteristics of the geometry, we can examine whether the Riemann and Ricci tensor components and the curvature scalars are finite across the entire coordinate domain. In the frame basis, the non-zero Riemann curvature components can be expressed as
\begin{equation}\label{eq59}
\begin{aligned}
     R^{0}{}_{110}&=-\frac{b_{0}^{2}(3r^{8}-12g^4r^4+g^8)}{(r^4+g^4)^3}\\
      R^{0}{}_{220}&= R^{0}{}_{330}=R^{2}{}_{112}= R^{3}{}_{113}=\frac{b_{0}^{2}(r^4-g^4)}{(r^4+g^4)^2}\\
     R^{3}{}_{223}&=-\frac{b_{0}^{2}}{r^4+g^4}
\end{aligned}    
\end{equation}
The non-zero Ricci tensor components in the frame basis are,
\begin{equation}\label{eq60}
\begin{aligned}
     R_{00}=-R_{11}&=-\frac{b_{0}^{2}(r^8-12g^4r^4+3g^8)}{(r^4+g^4)^3}\\
      R_{22}=R_{33}&=\frac{b_{0}^{2}(3g^4-r^4)}{(r^4+g^4)^2} 
\end{aligned}    
\end{equation}
Observe that as $r\to 0$, the values of each component of the Riemann and Ricci tensors are finite. When $r$ approaches infinity, all components converge to zero, guaranteeing that spacetime is asymptotically flat.

\noindent A spacetime is regular only if the curvature scalars are regular. In accordance with \cite{Kar}, we investigate three scalar invariants:

\noindent The Ricci scalar is given as
\begin{equation}\label{eq61}
    R=g^{\mu\nu}R_{\mu\nu}=\frac{4b_{0}^{2}(3g^{8}-5g^{4}r^{4})}{(r^{4}+g^{4})^{3}}
\end{equation}
The Ricci contraction becomes
\begin{equation}\label{eq62}
    R_{\mu\nu}R^{\mu\nu}=\frac{4b_{0}^{2}(r^{16}-14g^4r^{12}+74g^8r^8-30g^{12}r^4+9g^{16})}{(r^4+g^4)^6}
\end{equation}
Finally, the Kretschmann scalar turns out to be
\begin{equation}\label{eq63}    K=R_{\mu\nu\lambda\delta}R^{\mu\nu\lambda\delta}=\frac{8(3g^{16}-10g^{12}r^{4}+74g^{8}r^{8}-34g^{4}r^{12}+7r^{16})b_{0}^{4}}{(r^4+g^4)^6}
\end{equation}
Similar to the curvature tensors, all curvature scalars are also finite as $r\to 0$. This observation demonstrates the regular nature of the above geometry (\ref{eq58}).

\subsubsection{\textbf{Energy conditions}}
\noindent Let us now look at the energy conditions (restricted by GR) for the matter required to support the above geometry. One defines $\rho=-T_{t}^{t}$, $\tau=T_{r}^{r}$ and $p=T_{\theta}^{\theta}=T_{\phi}^{\phi}$. By using Eqs. (\ref{eq52}), (\ref{eq53}) and substituting the specific values of the model parameters provided in Eq. (\ref{eq57}), it is possible to express,
\begin{equation}\label{eq64}
    \begin{aligned}
        \rho=&-\tau=\frac{b_{0}^{2}(3g^4-r^4)}{\kappa(r^4+g^4)^2}\\
        p=&-\frac{b_{0}^{2}(3g^8-12g^4r^4+r^8)}{\kappa(r^4+g^4)^3}
    \end{aligned}
\end{equation}
The necessary conditions comprising the NEC (Null Energy Condition) are $\rho+\tau\geq0$ and $\rho+p\geq0$. The first condition is marginally satisfied since $\rho+\tau=0$. For the second condition, one has
\begin{equation}\label{eq65}
    \rho+p=-\frac{2b_{0}^{2}r^{4}(r^4-7g^4)}{\kappa(r^4+g^4)^3}
\end{equation}
Thus, the second NEC requirement is satisfied, but within a finite domain of the radial coordinate, i.e. if $r^4\leq7g^4$. Hence, NEC is not validated for all values of $r$. Showing the violation of NEC is sufficient to infer that other energy conditions, namely, weak, strong and dominant energy conditions, will also be violated \cite{Visser}.

\noindent Therefore, a regular spacetime is formed when the NLE model in Eq.(\ref{eq4}) is minimally coupled to GR via a magnetic monopole. As we have discussed previously, our Lagrangian eliminates the divergence in Maxwell's theory in flat spacetime. We have found that when coupled to gravity, we can resolve the spacetime singularity, too, for certain parameter values $(0<g^{2}\leq 0.5b_{0}^{2})$ in the presence of a magnetic monopole. Therefore, the assertion made in the Introduction on producing regular spacetimes is validated.

\subsection{The electric naked singularity solution}\label{VIB}
\noindent In the preceding section, we used a minimal coupling between our NLE model and Einstein's gravity and derived a family of regular spacetimes using a magnetic monopole source. In this subsection, we will replicate the exercise by examining an electric charge instead of a magnetic monopole, leading to the emergence of a naked singularity.   

\noindent It is essential to mention that for a spherically symmetric Schwarzschild-gauge geometry, the electric field of a point charge governed by a NLE Lagrangian is identical to what we get in flat spacetime background. Our assumption of spacetime in Eq.(\ref{eq50}) satisfies the above symmetries. Therefore, in our model, the electric field produced by a point charge in the `curved spacetime' (\ref{eq50}) can be represented by the expression in Eq.(\ref{eq16}). Here, we are looking for the purely electric solution, $F_{\theta\phi}=0$. Hence, the components of the energy-momentum tensor with the help of Eq.(\ref{eq51}) may be obtained as follows,
\begin{equation}\label{eq66}
    T_{t}^{t}=T_{r}^{r}=L(F)-2F L_{F}=\frac{\gamma E^{2}(4+3\eta E^{2}(8+\eta E^{2}))}{(-2+\eta E^{2})^{3}}
\end{equation}
and,
\begin{equation}\label{eq67}
    T_{\theta}^{\theta}=T_{\phi}^{\phi}=L(F)=\frac{\gamma E^{2}(2+3\eta E^{2})}{(-2+\eta E^{2})^{2}}
\end{equation}
where we substitute $F=-E^2/2$. The left-hand side of Einstein's equation is previously mentioned in Eq.(\ref{eq54}). Therefore, the differential equations for the metric function $f(r)$ obtained from Einstein's equation are as follows,
\begin{equation}\label{eq68}
\begin{aligned}
    \frac{f^{\prime}}{r}+\frac{f-1}{r^2}&=\frac{\gamma\kappa E^{2}(4+3\eta E^{2}(8+\eta E^{2}))}{(-2+\eta E^{2})^{3}} \\
    \frac{f^{\prime\prime}}{2}+\frac{f^{\prime}}{r}&=\frac{\gamma\kappa E^{2}(2+3\eta E^{2})}{(-2+\eta E^{2})^{2}}
\end{aligned}
\end{equation}
where the electric field $(E)$ is a function of $r$. As the electric field equation (\ref{eq16}) is a polynomial equation for $r$, it is difficult to solve the above two equations analytically. Nevertheless, in the weak field limit $(\eta F<<1)$, it is possible to get an analytic solution for the metric function $f(r)$ by treating it as a perturbation series in $\eta$. We write
\begin{equation}\label{eq69}
    f(r)=1+\frac{c}{r}+\eta^{0}f_{(0)}+\eta f_{(1)}+\eta^{2} f_{(2)}+O(\eta^{3})
\end{equation}
where $1+\frac{c}{r}$ is the known vacuum solution of Einstein's equation and $f_{(0)}$, $f_{(1)}$, $f_{(2)}$ are zeroth, first, second order corrections, respectively. There are other higher-order terms, but we choose to neglect them for now. By utilising the perturbation series of the electric field as represented in Eq.(\ref{eq22}), it is possible to write down the Einstein equations at various orders of $ \eta$. This gives,

\noindent Zeroth order:
\begin{equation}\label{eq70}
    \frac{f_{(0)}^{\prime\prime}}{2}+\frac{f_{(0)}^{\prime}}{r}=\frac{\gamma\kappa}{2}E_{(0)}^{2}
\end{equation}
First order:
\begin{equation}\label{eq71}
     \frac{f_{(1)}^{\prime\prime}}{2}+\frac{f_{(1)}^{\prime}}{r}=\frac{\gamma\kappa}{2}\left(2E_{(0)}E_{(1)}+\frac{5}{2}E_{(0)}^{4}\right)
\end{equation}
Second-order:
\begin{equation}\label{eq72}
     \frac{f_{(2)}^{\prime\prime}}{2}+\frac{f_{(2)}^{\prime}}{r}=\frac{\gamma\kappa}{2}\left(E_{(1)}^{2}+10 E_{(0)}^{3}E_{(1)}+2E_{(0)}E_{(2)}+\frac{9}{4}E_{(0)}^{6}\right)
\end{equation}
We have utilized the second differential equation in the Eq.(\ref{eq68}). The first one is expected to satisfy any solution of the second equation trivially. Thus, the different order equations mentioned above can be solved using the values of $E_{(0)}$, $E_{(1)}$, and $E_{(2)}$ from the Eq.(\ref{eq22}). This will yield the following metric function,
\begin{equation}\label{eq73}
    f(r)=1+\frac{c}{r}+c_{(0)}\eta^0\frac{e^{2}}{r^{2}}+c_{(1)}\eta\frac{e^{4}}{r^{8}}+c_{(2)}\eta^{2}\frac{e^{6}}{r^{10}}+O(\eta^{3})
\end{equation}
where $c_{(0)}$, $c_{(1)}$ and $c_{(2)}$ are some arbitrary constants (functions of $\gamma$ and $\kappa$). It is evident that the above metric function represents the corrections made to a Reissner-Nordstr\"om like metric. Nevertheless, the fundamental structure of the spacetime remains ambiguous in this formulation. To improve on this, we numerically solve Einstein's equation (\ref{eq68}) for any arbitrary value of $\eta$. The method consists of the following steps:

\noindent Let us take into consideration the first differential equation of (\ref{eq68}) with the following modification of the variable $E^2\to\Tilde{E}$. Therefore, we obtain,
\begin{equation}\label{eq74}
    r\frac{df(r)}{dr}+f(r)-1=r^2\frac{\gamma\kappa \Tilde{E}(4+3\eta \Tilde{E}(8+\eta \Tilde{E}))}{(-2+\eta \Tilde{E})^{3}} 
\end{equation}
Next, one can perform another necessary transformation of the metric function $f(r) \to f(\Tilde{E})$. In other words, the variable $\Tilde{E}$ can be used as a substitute for $r$, based on the relation of $r$ and $\Tilde{E}$ described in Eq.(\ref{eq16}). As a result, the equation above becomes,
\begin{equation}\label{eq75}
    \frac{df}{d\Tilde{E}}\left(r\frac{d\Tilde{E}}{dr}\right)+f(\Tilde{E})-1=r^{2}\frac{\gamma\kappa \Tilde{E}(4+3\eta \Tilde{E}(8+\eta \Tilde{E}))}{(-2+\eta \Tilde{E})^{3}} 
\end{equation}
As $r\in[0,\infty)$, $\Tilde{E}$ ranges between $[2/\eta,0]$ which is also evident from Eq.(\ref{eq16}). After that, substituting $d\Tilde{E}/dr$ in the above Eq.(\ref{eq75}) we have,
\begin{equation}\label{eq76}
    \frac{4\Tilde{E}(-2+\eta \Tilde{E})(2+7\eta\Tilde{E})}{4+\eta\Tilde{E}(52+21\eta\Tilde{E})}\frac{df}{d\Tilde{E}}+f(\Tilde{E})-1=-\frac{e\kappa\sqrt{\Tilde{E}}(4+3\eta\Tilde{E}(8+\eta\Tilde{E}))}{16\pi(2+7\eta\Tilde{E})}
\end{equation}
From Eq.(\ref{eq76}), it is now possible to find $f(\Tilde{E})$ numerically for a given value of $\eta$. To summarize, we have $f(\Tilde{E})$ and $\Tilde{E}(r)$ from Eq. (\ref{eq16}). This allows us to evaluate the behaviour of the metric function $f(r)$ as a function of the radial coordinate $r$ by numerically inverting the abovementioned functions.

\begin{figure}[h]
\centering
\includegraphics[width=0.6\textwidth]{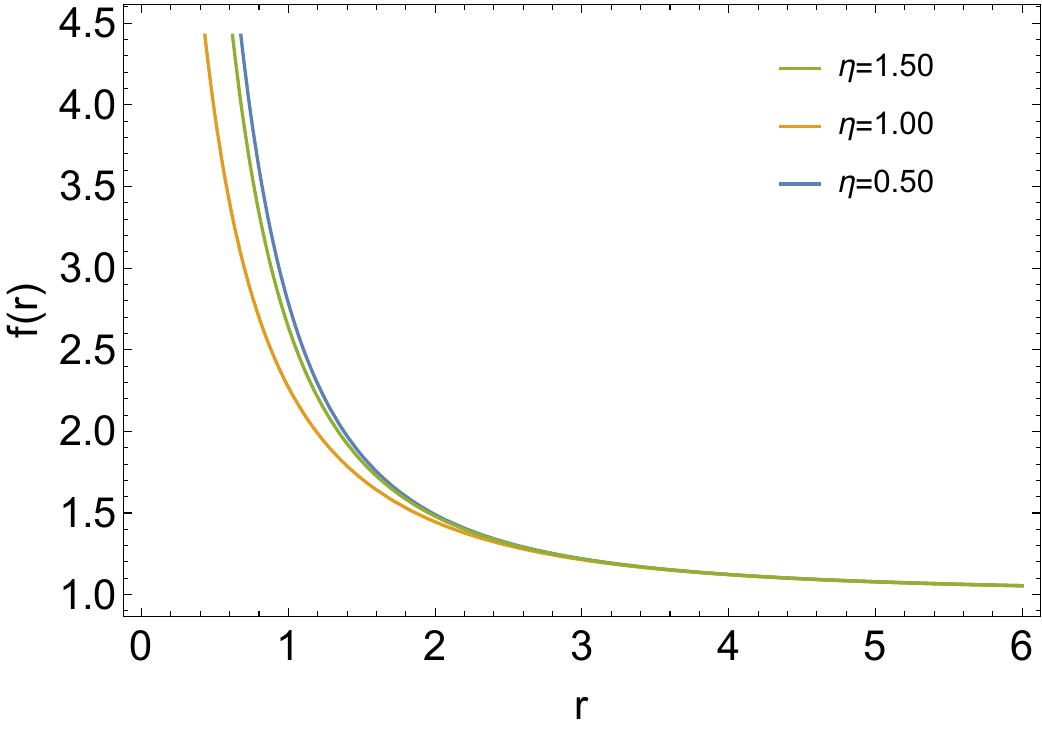}
\caption{Plot of metric function $f(r)$ with $r$ for different values of $\eta$, assuming $e=2$ and $\gamma=1/2\pi$}
\label{fig:nakedsingularity}
\end{figure}
\noindent Figure \ref{fig:nakedsingularity} illustrates the variation of the metric function with respect to the radial coordinate. It is clear that the function $f(r)$ approaches unity asymptotically, indicating that the metric is asymptotically flat. Additionally, $f(r)$ does not cross the line $f(r)=0$, meaning there is no horizon-like structure. To determine whether the spacetime is a naked singularity, we need to analyze the functional behaviour of the curvature scalars across all values of the radial coordinate.

\noindent In a spherically symmetric Schwarzschild-gauge metric, the independent scalar invariants can be expressed as a function of the metric function and its radial derivative as,

\noindent Ricci scalar:
\begin{equation}\label{eq77}
    R=g_{\mu\nu}R^{\mu\nu}=\frac{4}{r^2}-\frac{4f(r)}{r^2}-\frac{8f^{\prime}(r)}{r}-2f^{\prime\prime}(r)
\end{equation}
with the help of equation (\ref{eq68}), one can convert the above equation in terms of $\Tilde{E}$
\begin{equation}\label{eq78}
    R=-8\kappa\left(L-FL_{F}\right)=-\frac{8\gamma\kappa\eta\Tilde{E}^{2}\left(10+3\eta\Tilde{E}\right)}{\left(-2+\eta\Tilde{E}\right)^{3}}
\end{equation}
The Ricci contraction is given as,
\begin{equation}\label{eq79}
    R_{\mu\nu}R^{\mu\nu}=8\left(\frac{f^{\prime}(r)}{r}+\frac{f(r)-1}{r^2}\right)^2+8\left(\frac{f^{\prime}(r)}{r}+\frac{f^{\prime\prime}(r)}{2}\right)^2
\end{equation}
Again, in terms of $\Tilde{E}$, Eq.(\ref{eq79}) becomes
\begin{equation}\label{eq80}
    \begin{aligned}
    R_{\mu\nu}R^{\mu\nu}&=8\kappa\left((L-2FL_{F})^2+L^{2}\right)\\
    &=\frac{16\kappa\gamma^{2}\Tilde{E}^{2}\Big\{16+\eta\Tilde{E}\left(112+\eta\Tilde{E}\left(296+3\eta\Tilde{E}\left(20+3\eta\Tilde{E}\right)\right)\right)\Big\}}{\left(-2+\eta\Tilde{E}\right)^{6}}
\end{aligned}
\end{equation}
It is evident from the expressions of the scalars that at $\Tilde{E}=2/\eta$ or, in terms of radial coordinate, at $r=0$, the scalars diverge, indicating a physical spacetime singularity at $r=0$. One can evaluate the Kretschmann scalar $(R_{\mu\nu\lambda\delta}R^{\mu\nu\lambda\delta})$ which also shows a divergence at $r=0$ (not shown here). Therefore, a nonlinear electric monopole governed by the Lagrangian (\ref{eq4}), minimally coupled to Einstein's gravity, leads to a `naked singularity'.

\section{Conclusion}\label{VII}
\noindent We now summarise our research with concluding remarks.
\begin{enumerate}
    \item In this study, we address the inconsistency in the electric field at its location as well as the infinite self-energy of a point charge in Maxwell's electrodynamics and propose a novel NLE Lagrangian. We first examine our model in flat spacetime and provide a comprehensive analysis of the field equation, focusing specifically on the electrostatic limit. It has been shown that the electric field of a point charge reaches a maximum value, and its self-energy is finite.
    \item We demonstrate how a plane electromagnetic wave displays vacuum birefringence in the presence of a uniform background electric field. It has been argued that when an elementary excitation propagates across spacetime, the principles of causality and unitarity are maintained for all values of the electric field but in a restricted domain of the magnetic field. 
    \item The coupling of Einstein's GR with our NLE model leads to the formation of various spacetime solutions, which depend on the characteristics of the source field. For instance, a nonlinear magnetic monopole yields a family of regular solutions. These solutions can represent double-horizon and single-horizon regular black holes or a family of regular spacetimes without a horizon, depending on specific parameter values. However, if the source is an electric field caused by a point charge, we encounter naked singularities. Further, the non-singular and singular character of regular black holes and naked singularities, respectively, are confirmed by analyzing scalar invariants.
\end{enumerate}
Although we claim that the NLE model (discussed in this article) is a viable source for generating different spacetimes (when coupled to GR), the causality
conditions do not hold for all values of the magnetic field. Only for a restricted domain of the magnetic field, the causality and unitarity conditions hold. Moreover, even though the matter Lagrangian does not contain any fractional power of $F$, yet the required matter must violate the energy conditions (assuming GR field equations) in order to give rise to regular black holes. And when our model is coupled to GR, there always seems to be an unresolved 
singularity either in the source field or in the geometry. Below, we have outlined our achievements and limitations in tabular form.
\begin{center}
\begin{tabular}{|c|c| } 
\hline
Source field & Spacetime solution\\
\hline
\hline
Magnetic monopole (\textit{singular}) & Regular black hole (\textit{non-singular}) \\ 
\hline
Electric charge (\textit{non-singular}) & Naked singularity (\textit{singular}) \\
\hline
\end{tabular}
\end{center}
Therefore, it may be inferred that if the coupling between our Lagrangian and GR results in a regular solution, then the source field is inherently singular, or \emph{vice versa}. However, since physical quantities (source fields or others) are defined within the framework of spacetime, spacetime singularities pose a more fundamental challenge than field singularities. Nevertheless, neither can be ignored. Therefore, a simultaneous resolution of spacetime and field singularities, which is not achieved here, will be a remarkable accomplishment and is our future goal. In conclusion, we have a novel NLE Lagrangian model that effectively addresses the issue of field divergence
(for a point charge) in Maxwell's electrodynamics in flat spacetime and
when coupled to gravity, can resolve (with singular matter fields) the spacetime singularity in classical GR. 

\noindent Our research can be extended to explore the gravitational coupling of our generic Lagrangian (\ref{n1}) with other parameter values. Moreover, it is possible to study the solution of Einstein's equation coupled to our specific NLE model (\ref{eq4}) when the source field is a dyonic configuration. In addition, the regular black hole discussed in this article has a well-behaved functional form, making it appropriate for a variety of astrophysical studies. The shadow and quasinormal modes of the regular spacetime may be computed, and their feasibility can be tested with observations.

\section*{Acknowledgements}
\noindent AK expresses gratitude to Prof. Sayan Kar for his continuous inspiration and guidance throughout the successful completion of this work. Additionally, AK acknowledges Prof. Kar's thorough reading of the manuscript. He also thanks Pritam Banerjee and Aditya Sharma for some useful discussions.


\begin{thebibliography}{Rubinsteinetal}

\bibitem{Born1} M. Born, \href{https://doi.org/10.1098/rspa.1934.0010}{Proc. R. Soc. A \textbf{143}, 410 (1934)}.

\bibitem{Born2}  M. Born and L. Infeld, \href{https://doi.org/10.1098/rspa.1934.0010}{Proc. R. Soc. A \textbf{144}, 425 (1934)}. 

\bibitem{Heisenberg} W. Heisenberg and H. Euler, \href{ https://doi.org/10.1007/BF01343663}{Z. Phys. \textbf{98}, 714 (1936)}

\bibitem{Townsend} I. Bandos, K. Lechner, D. Sorokin, and P. Townsend, \href{https://doi.org/10.1103/PhysRevD.102.121703}{Phys. Rev. D \textbf{102}, 121703 (2020)}

\bibitem{Soleng} H.H Soleng, \href{https://doi.org/10.1103/PhysRevD.52.6178}{Phys. Rev. D \textbf{52}, 6178 (1995)}

\bibitem{Kruglov1} S.I Kruglov, \href{https://doi.org/10.1142/S0218271816400022}{Internat. J. Modern Phys. D \textbf{25}, 11 (2016)}

\bibitem{Kruglov2} S.I Kruglov, \href{https://doi.org/10.1016/j.physleta.2023.129248}{Phys. Lett. A \textbf{493}, 129248 (2024)}

\bibitem{Kruglov3} S.I Kruglov, \href{https://doi.org/10.1002/andp.201500142}{Annalen Der Physik \textbf{527}, 397-401 (2015)}

\bibitem{Kruglov6} S.I Kruglov, \href{https://doi.org/10.1006/aphy.2001.6186}{Ann. of Phys. \textbf{293}, 228 (2001)}

\bibitem{Kruglov7} S.I Kruglov, \href{https://doi.org/10.1142/S0217732308026339}{Mod. Phys. Lett. A \textbf{23}, 245 (2008)}

\bibitem{Gaete} P. Gaete and J. Helay\"el-Neto, \href{https://doi.org/10.1140/epjc/s10052-014-2816-4}{ Eur. Phys. J. C \textbf{74}, 2816 (2014)} 

\bibitem{Hendi}  S. H. Hendi, \href{https://doi.org/10.1016/j.aop.2013.03.008}{Ann. of Phys. \textbf{333}, 282 (2013)}

\bibitem{Gullu} I. Gullu and S. H. Mazharimousavi, \href{https://doi.org/10.1088/1402-4896/abe498}{Phys. Scr. \textbf{96}, 045217 (2021)}

\bibitem{Balart} L. Balart, S. B. Herrera, G. Panotopoulos and A. Rincon, \href{https://doi.org/10.1016/j.aop.2023.169329}{Ann. of Phys. \textbf{454}, 169329 (2023)}

\bibitem{Mazharimousav} S. H. Mazharimousavi and M. Halilsoy,  \href{https://doi.org/10.1016/j.aop.2021.168579}{Ann. of Phys. \textbf{433}, 168579 (2021)}

\bibitem{Hawking1} S. Hawking and R. Penrose, \href{https://doi.org/10.1098/rspa.1970.0021}{Proc. Roy. Soc. Lon. A \textbf{314}, 529 (1970)}

\bibitem{Penrose} R. Penrose, \href{https://doi.org/10.1103/PhysRevLett.14.57}{Phys. Rev. Lett. \textbf{14}, 57 (1965)}

\bibitem{Hawking2} S. Hawking, \href{https://doi.org/10.1103/PhysRevLett.17.444}{Phys. Rev. Lett. \textbf{17}, 444 (1966)}

\bibitem{Senovilla} J. M. M. Senovilla and D. Garfinkle \href{https://doi.org/10.1088/0264-9381/32/12/124008}{Class. Quantum Grav. \textbf{32}, 124008 (2015)}

\bibitem{Crowther} K. Crowther and S. De Haro, \href{https://doi.org/10.48550/arXiv.2112.08531}{, arXiv:2112.08531 [gr-qc] (2021)}

\bibitem{Lifshitz} V. B. Berestetskii, E. M. Lifshitz, L. P. Pitaevskii, {\em `Course of Theoretical Physics, Volume 4: Quantum Electrodynamics (1982)'}. \href{https://doi.org/10.1016/C2009-0-24486-2}{ISBN 978-0-7506-3371-0}.

\bibitem{Sorokin} I. Bandos, K. Lechner, D. Sorokin, and P. K. Townsend, \href{https://doi.org/10.1007/JHEP03(2021)022}{J. High Energ. Phys. \textbf{03} (2021) 022}.

\bibitem{Fradkin} E.S. Fradkin, Arkady A. Tseytlin, \href{https://doi.org/10.1016/0370-2693(85)90205-9}{Phys.Lett.B \textbf{163} (1985) 123-130}

\bibitem{Ayon1} E. Ayon-Beato and A. Garcia, \href{https://doi.org/10.1103/PhysRevLett.80.5056}{Phys. Rev. Lett. \textbf{80}, 5056 (1998).}

\bibitem{Ayon2}  E. Ayon-Beato and A. Garcia, \href{https://doi.org/10.1016/S0370-2693%2899%2901038-2}{Phys. Lett. B \textbf{464}, 25 (1999).}

\bibitem{Bardeen} J. M. Bardeen, {\em in Proceedings of International Conference GR5, Tbilisi, USSR, 1968}, p. 174.

\bibitem{Hayward} S. A. Hayward, \href{https://link.aps.org/doi/10.1103/PhysRevLett.96.031103}{Phys. Rev. Lett. \textbf{96}, 031103 (2006).}

\bibitem{Roman} T.A. Roman and P.G. Bergmann, \href{https://link.aps.org/doi/10.1103/PhysRevD.28.1265}{Phys. Rev. D \textbf{28} 1265 (1983)}.

\bibitem{Dymnikova1} I. Dymnikova, \href{https://ui.adsabs.harvard.edu/link_gateway/1992GReGr..24..235D/doi:10.1007/BF00760226}{Gen. Relativ. Gravit. \textbf{24}, 235 (1992).}

\bibitem{Dymnikova2} I. Dymnikova, \href{https://doi.org/10.1142/S021827180300358X}{Int. J. Mod. Phys. D \textbf{12}, 1015(2003).}

\bibitem{Bronnikov3} K. A. Bronnikov, \href{https://link.aps.org/doi/10.1103/PhysRevD.63.044005}
{Phys. Rev. D \textbf{63}, 044005 (2001).}

\bibitem{Carballo2} R. Carballo-Rubio, F. Di Filippo, S. Liberati, C. Pacilio and M. Visser,  \href{https://doi.org/10.1007/JHEP07(2018)023}{J. High Energ. Phys. \textbf{07}, 023 (2018)}.

\bibitem{Bambi} C. Bambi and L. Modesto, \href{https://doi.org/10.1016/j.physletb.2013.03.025}{Phys. Lett. B \textbf{721}, 329 (2013).}

\bibitem{Carballo3} R. Carballo-Rubio, F. Di Filippo, S. Liberati and M. Visser,  \href{https://link.aps.org/doi/10.1103/PhysRevD.98.124009}{Phys. Rev. D \textbf{98}, 124009 (2018)}. 

\bibitem{Frolov3}V.P. Frolov and A. Zelnikov, \href{https://link.aps.org/doi/10.1103/PhysRevD.95.124028}{Phys. Rev. D \textbf{95}, 124028 (2017).}

\bibitem{Balart1} L. Balart and E. C. Vagenas, \href{https://link.aps.org/doi/10.1103/PhysRevD.90.124045}
{Phys. Rev. D \textbf{90}, 124045 (2014).}

\bibitem{Frolov} V.P. Frolov, \href{https://link.aps.org/doi/10.1103/PhysRevD.94.104056}
{Phys. Rev. D \textbf{94}, 104056 (2016).}

\bibitem{Ayon3} E. Ayon-Beato and A. Garcia, \href{https://doi.org/10.1023/A:1026640911319}{Gen. Relativ. Gravit. \textbf{31}, 629 (1999).}

\bibitem{Ayon4} E. Ayon-Beato and A. Garcia, \href{https://doi.org/10.1016/S0370-2693%2800%2901125-4}{Phys. Lett. B \textbf{493}, 149 (2000).}

\bibitem{Ayon5} E. Ayon-Beato and A. Garcia, \href{https://doi.org/10.1007/s10714-005-0050-y}{Gen. Relativ. Gravit. \textbf{37}, 635 (2005).}

\bibitem{Fan1} Z. Y. Fan, \href{https://doi.org/10.1140/epjc/s10052-017-4830-9}{Eur. Phys. J. C \textbf{77}, 266 (2017).}

\bibitem{Bronnikov4} K. A. Bronnikov, \href {https://link.aps.org/doi/10.1103/PhysRevLett.85.4641}{Phys. Rev. Lett. \textbf{85}, 4641 (2000)}

\bibitem{Bronnikov5}K. A. Bronnikov, \href{https://doi.org/10.1142/S0218271818410055}{Int. J. Mod. Phys. D \textbf{27}, 1841005 (2018)}

\bibitem{Smolic} A. Bokuli\'c, I. Smoli\'c, and T. Juri\'c, \href{https://doi.org/10.1103/PhysRevD.106.064020}{Phys. Rev. D \textbf{106}, 064020 (2022)}

\bibitem{Kolar} F. Di Filippo, I. Kol\'a\~r, D. Kubiznak, \href{https://doi.org/10.48550/arXiv.2404.07058}{arXiv:2404.07058 [gr-qc]}

\bibitem{Bueno} P. Bueno, P. A. Cano, R. A. Hennigar, \href{https://doi.org/10.48550/arXiv.2403.04827}{arXiv:2403.04827 [gr-qc]}

\bibitem{Bokulic} A. Bokuli\'c, E. Franzin, T. Juri\'c, I. Smoli\'c, \href{https://doi.org/10.1016/j.physletb.2024.138750}{Phys. Lett. B \textbf{854}, 138750 (2024)}

\bibitem{Bonanno} A. Bonanno, D. Malafarina, and A. Panassiti, \href{https://doi.org/10.1103/PhysRevLett.132.031401}{Phys. Rev. Lett. \textbf{132}, 031401 (2024)}

\bibitem{Lessa} L. A. Lessa, J. E. G. Silva, \href{https://doi.org/10.48550/arXiv.2305.18254}{arXiv:2305.18254 [gr-qc]}

\bibitem{Estrada} M. Estrada, R. Aros, \href{https://doi.org/10.1016/j.physletb.2023.138090}{Phys.Lett.B \textbf{844}, 138090 (2023)}

\bibitem{Karakasis} T. Karakasis, N. E. Mavromatos, E. Papantonopoulos, \href{https://doi.org/10.1103/PhysRevD.108.024001}{Phys. Rev. D \textbf{108}, 024001 (2023)}

\bibitem{Ovalle} J. Ovalle, R. Casadio, A. Giusti, \href{https://doi.org/10.1016/j.physletb.2023.138085}{Phys.Lett.B \textbf{844}, 138085 (2023)}

\bibitem{Ara} A. A. Ara\'ujo Filho, \href{https://doi.org/10.1088/1361-6382/ad0a19}{Class. Quant. Grav., \textbf{41}, 015003 (2023)}

\bibitem{Fan} Z. Y. Fan and X. Wang, \href{https://doi.org/10.1103/PhysRevD.94.124027}{Phys. Rev. D \textbf{94}, 124027 (2016)}.

\bibitem{Toshmatov} B. Toshmatov, Z. Stuchlík, and B. Ahmedov, \href{https://doi.org/10.1103/PhysRevD.98.028501}{Phys. Rev. D \textbf{98}, 028501 (2018)}

\bibitem{Bronnikov1} K. A. Bronnikov, \href{https://doi.org/10.1103/PhysRevD.96.128501}{Phys. Rev. D \textbf{96}, 128501 (2017)}.

\bibitem{Bronnikov2} K. A. Bronnikov, \href{https://doi.org/10.1103/PhysRevD.101.128501}{Phys. Rev. D \textbf{101}, 128501 (2020)}.

\bibitem{Marco} Marco A.A. de Paula, Haroldo C.D. Lima Junior, Pedro V.P. Cunha, Luís C.B. Crispino, \href{https://doi.org/10.1103/PhysRevD.108.084029}{Phys.Rev.D \textbf{108}, 084029 (2023)}

\bibitem{Krug} S. Fernando, D. Krug, \href{https://doi.org/10.1023/A:1021315214180}{Gen. Relativity Gravitation \textbf{35}, 129 (2003)}

\bibitem{Dey} T. K. Dey \href{https://doi.org/10.1016/j.physletb.2004.06.047}{Phys. Lett. B \textbf{595}, 484 (2004)}

\bibitem{Diaz} J. Diaz-Alonso, D. Rubiera-Garcia, \href{https://doi.org/10.1103/PhysRevD.81.064021}{Phys. Rev. D \textbf{81}, 064021 (2010)}

\bibitem{Aiello} M. Aiello, R. Ferraro, G. Giribet, \href{https://doi.org/10.1103/PhysRevD.70.104014}{Phys. Rev. D \textbf{70}, 104014 (2004)}

\bibitem{Paul} P. Paul and S. I. Kruglov, \href{https://doi.org/10.48550/arXiv.2403.02056}{arXiv:2403.02056 [gr-qc]}

\bibitem{Junior} José Tarciso S. S. Junior, Francisco S. N. Lobo and Manuel E. Rodrigues, \href{https://doi.org/10.1140/epjc/s10052-024-12696-8}{Eur. Phys. J. C \textbf{84}, 332 (2024)}

\bibitem{Habibina} A S. Habibina, B.N. Jayawiguna, H.S. Ramadhan, \href{https://doi.org/10.1007/s10714-021-02882-4}{Gen.Rel.Grav. \textbf{53}, 113 (2021)}

\bibitem{Gibbons} G. W. Gibbons and D. Rasheed, \href{https://doi.org/10.1016/0550-3213%2895%2900409-L}{Nucl. Phys., B \textbf{454}, 185 (1995)}

\bibitem{Blaschke} D.N. Blaschke, F. Gieres, M. Reboud, M. Schweda, \href{https://doi.org/10.1016/j.nuclphysb.2016.07.001}{ Nucl. Phys. B \textbf{912} (2016) p. 192–223}

\bibitem{Belinfante} F. J. Belinfante, \href{https://doi.org/10.1016/S0031-8914(40)90091-X}{Physica \textbf{7}, 5 (1940) p. 449-474.}

\bibitem{Novello2} M. Novello, S. E. Perez Bergliaffa and J. M. Salim, \href{https://doi.org/10.1088/0264-9381/17/18/316}{Class. Quant. Grav. \textbf{17}, 3821-3832 (2000)} 

\bibitem{Novello} M. Novello, V. A. De Lorenci, J. M. Salim, and R. Klippert,  \href{https://doi.org/10.1103/PhysRevD.61.045001}{Phys. Rev. D \textbf{61}, 045001 (2000)}

\bibitem{Adler} S. L. Adler, \href{https://doi.org/10.1088/1751-8113/40/5/F01}{J. Phys. A \textbf{40}, F143 (2007)}

\bibitem{Biswas} S. Biswas and K. Melnikov, \href{https://doi.org/10.1103/PhysRevD.75.053003}{Phys. Rev. D \textbf{75}, 053003 (2007)}


\bibitem{Tomizawa} S. Tomizawa and R. Suzuki, \href{https://doi.org/10.1103/PhysRevD.108.124072}{Phys. Rev. D \textbf{108}, 124072 (2023)}

\bibitem{Schellstede} G. O. Schellstede, V. Perlick, C. L\"ammerzahl, \href{https://doi.org/10.1002/andp.201600124}{Ann. of Phys. \textbf{528}, 738 (2016)}

\bibitem{Melo} C. A. M. de Melo, L. G. Medeiros, and P. J. Pompei \href{https://doi.org/10.1142/S021773231550025X}{Mod. Phys. Lett. A \textbf{30}, 1550025 (2015)}

\bibitem{Russo} J. G. Russo, P. K. Townsend, \href{https://doi.org/10.48550/arXiv.2401.04167}{arXiv:2401.04167 [hep-th]}

\bibitem{Russo2} J. G. Russo, P. K. Townsend, \href{https://doi.org/10.1007/JHEP01%282023%29039}{JHEP \textbf{01}, 039 (2023)}

\bibitem{Kruglov4} S. I. Kruglov, \href{https://doi.org/10.1088/1751-8113/43/37/375402}{J. Phys. A \textbf{43}, 375402 (2010)}

\bibitem{Kruglov5} S. I. Kruglov, \href{https://doi.org/10.1103/PhysRevD.75.117301}{Phys. Rev. D \textbf{75}, 117301 (2007)}

\bibitem{Shabad} A. E. Shabad, V. V. Usov, \href{https://doi.org/10.1103/PhysRevD.83.105006}{Phys. Rev. D \textbf{83}, 105006 (2011)}

\bibitem{Rosen} A. Einstein and N. Rosen, \href{https://link.aps.org/doi/10.1103/PhysRev.48.73}{Phys. Rev. \textbf{48}, 73 (1935).}

\bibitem{Kar} A. Kar and S. Kar, \href{https://doi.org/10.1007/s10714-024-03238-4}{Gen. Relativ. Gravit. \textbf{56}, 52 (2024)}

\bibitem{Visser} M. Visser, {\em `Lorentzian wormholes: From Einstein to Hawking'},
(AIP Press, now Springer, New York, 1995).
\end{thebibliography}
\end{document}